\documentclass[twocolumn,showpacs,preprintnumbers,amsmath,amssymb]{revtex4}

\usepackage{graphicx}
\usepackage{rotating}

\begin{document}

\draft

\title{Tuning the scattering length with an optically induced Feshbach resonance}

\author{M. Theis,$^{1}$ G. Thalhammer,$^{1}$ K.
Winkler,$^{1}$ M. Hellwig,$^{1}$ G. Ruff,$^{1,\dagger}$
 R. Grimm,$^{1,2}$ and J. Hecker Denschlag$^{1}$}

\address{$^{1}$Institut f\"ur Experimentalphysik, Universit\"at Innsbruck,
 Technikerstra{\ss}e 25, 6020 Innsbruck,
Austria\\$^{2}$Institut f\"ur Quantenoptik und Quanteninformation,
\"Osterreichische Akademie der Wissenschaften, 6020 Innsbruck,
Austria}

\date{\today}


\begin{abstract}
We demonstrate optical tuning of the scattering length in a
Bose-Einstein condensate as predicted by Fedichev {\em et al.}
[Phys. Rev. Lett. {\bf 77}, 2913 (1996)]. In our experiment atoms
in a $^{87}$Rb condensate are exposed to laser light which is
tuned close to the transition frequency to an excited molecular
state. By controlling the power and detuning of the laser beam we
can change the atomic scattering length over a wide range. In view
of laser-driven atomic losses we use Bragg spectroscopy as a fast
method to measure the scattering length of the atoms.
\end{abstract}

\maketitle

\narrowtext

 The great progress in the field of ultracold quantum gases
 in recent years
can be largely attributed to the existence of magnetically tunable
Feshbach resonances \cite{Tiesinga}. Since their first
experimental introduction into the field
\cite{Inouye,Courteille,Roberts} they have been widely used to
arbitrarily tune the interactions between atoms. A plethora of
experiments has been performed ranging for example from ultra-high
resolution molecular spectroscopy \cite{Chin} to the creation of
bright matter wave solitons \cite{Kaykovich}
 as well as the
production of new atomic \cite{Cornish} and molecular
\cite{Jochim} Bose-Einstein condensates (BEC).

In general a Feshbach  resonance occurs when a
colliding pair of atoms
 is resonantly coupled to a molecular bound state.
 A magnetically tunable Feshbach resonance is based on Zeeman
shifting a bound molecular state into resonance with the
scattering state.
 Alternative coupling schemes for inducing Feshbach resonances
 have been proposed but never experimentally applied to control
  atomic interactions. The use of  radiofrequency
   \cite{Moerdijk} and static electric fields
\cite{Marinescu} was suggested.   Fedichev {\em et
al.}~\cite{Fedichev} proposed optical coupling of the scattering
state with the molecular state which was theoretically analyzed
further in \cite{Bohn,Kokoouline}. This scheme, often referred to
as ``optical Feshbach resonance'', can be controlled via laser
detuning and laser power.

 Inducing Feshbach resonances with
optical fields offers experimental advantages compared to magnetic
fields. The intensity and detuning of optical fields can be
rapidly changed. Furthermore complex spatial intensity
distributions can be easily produced which result in corresponding
scattering length patterns across the sample. Optical transitions
are always available even when no magnetic Feshbach resonances
exist. Recently Fatemi {\em et al.} \cite{Fatemi} observed optical
Feshbach resonances in photoassociation spectroscopy.  They used
photoionization to probe optically induced changes in the
scattering wave function. However, the direct influence of the
optical  Feshbach resonance on the atomic scattering properties
was not studied.

In this Letter we report a direct measurement of the atomic
scattering length $a$ in a BEC of $^{87}$Rb $| F = 1, m_F = -1
\rangle$ as we cross an optical Feshbach resonance. With moderate
laser intensities of about 500 W/cm$^2$ we can change
 the scattering length over one order of magnitude from
 10 a$_0$ to 190 a$_0$
 (a$_0$ = 1 Bohr
radius).
\begin{figure}
\includegraphics[width=3in]{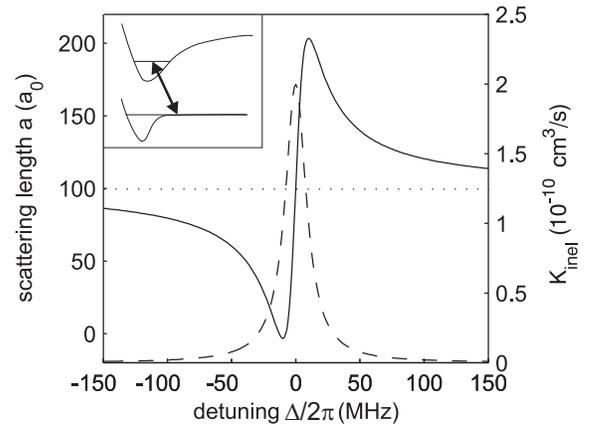}
\caption{Scattering length $a$ (solid line) and inelastic
collision rate coeffient K$_\text{inel}$ (broken line) as a
function of the laser detuning from the photoassociation
resonance. The curves are based on
Eqs.~(\ref{eq:scatt_length_Julienne}) and
(\ref{eq:scatt_length_Julienne2}) for typical experimental
parameters: $\Gamma_\text{stim}/2\pi= 54\text{ kHz}$,
$\Gamma_\text{spon}/2\pi= 20\text{ MHz}$, $k_{i}= 2.47\times
10^{5}\, \text{m}^{-1}$, $a_\text{bg} = 100$ a$_0$ (dotted line).
Inset: Scheme for optically coupling the scattering state with an
excited molecular state.} \label{theoryplot}
\end{figure}

To optically modify the scattering length we use laser light tuned
close to a photoassociation resonance which couples the continuum
state of incoming free atoms to an excited molecular level (see
inset in Fig.~\ref{theoryplot}). This changes the wavefunction and
consequently the scattering length of the scattering state. It
also leads to atomic loss due to spontaneous decay via the
molecular state.  The resonant transition rate between the
continuum state and the molecular state, which we denote
$\Gamma_\text{stim}$, is proportional to the laser intensity. In
our experiment $\Gamma_\text{stim}/2\pi$ is on the order of a few
10 kHz. This is three orders of magnitude less than the
spontaneous decay rate $\Gamma_\text{spon}$ from the excited
molecular state. In \cite{Bohn} Bohn and Julienne give convenient
expressions for the scattering length $a$ and the inelastic
collision rate coefficient $K_\text{inel}$ which describes the
photoassociation loss. For $\Gamma_\text{stim} \ll
\Gamma_\text{spon}$ these expressions can be approximated and, for
a condensate \footnote{For condensed atoms the collision rate
coefficient is only half of the coefficient for thermal atoms as
all atoms share the same quantum state.}, read:
\begin{eqnarray}
  \label{eq:scatt_length_Julienne}
  a = a_\text{bg} + \frac{1}{2 k_i} \
  \frac{ \Gamma_\text{stim} \ \Delta}
       {\Delta^2 + (\Gamma_\text{spon}/2)^2}
\\ \label{eq:scatt_length_Julienne2}
  K_\text{inel} = \frac{2\pi\hbar}{ m} \ \frac{1}{k_i} \
   \frac{\Gamma_\text{stim} \ \Gamma_\text{spon}}
                     {\Delta^2 + (\Gamma_\text{spon}/2)^2}
\end{eqnarray}
where $a_\text{bg}$ is the scattering length in the absence of
light, $\Delta$ is the detuning from the photoassociation line,
$m$ the atomic mass and $\hbar k_i$ the relative momentum of the
collision. Fig.~\ref{theoryplot} shows $a$ and $K_\text{inel}$ as
a function of the detuning $\Delta$ for typical experimental
parameters. According to Eqs.~(\ref{eq:scatt_length_Julienne}) and
(\ref{eq:scatt_length_Julienne2}) one should in general choose
large detuning and strong coupling in order to maximize the change
in scattering length while keeping the losses low.

Our experiments are carried out with an almost
 pure $^\text{87}$Rb
condensate in the $| F=1, m_F = -1 \rangle$ spin state with
typically $1\times 10^6$ atoms. The setup uses a magnetic
transport scheme \cite{Greiner} to transfer atoms from a
magneto-optical trap (MOT) chamber to a glass cell where the BEC
is produced by rf-evaporation in a cigar shaped QUIC trap
\cite{QUIC} with trap frequencies $\omega_\text{axial}/2\pi =
15\text{ Hz}$ and $\omega_\text{radial}/2\pi = 150\text{ Hz}$
\footnote{Details of our setup will be described elsewhere.}.
 The intensity stabilized
photoassociation laser beam ($\approx 40$ mW) is derived from a
Ti:Sa laser. It is aligned along the axial direction of the cigar
shaped BEC and has a waist radius of 76 $\mu$m. Its linear
polarization is perpendicular to the trapping magnetic bias field
of 2 Gauss. In our experiments we limit the maximum laser
intensities to about 500 W/cm$^2$ because we observe the
appearance of a growing component of thermal atoms for higher
intensities \footnote{We attribute this heating to the appearance
of an uncontrolled corrugation in the photoassociation laser beam
which transfers momentum components to the condensate atoms. The
corrugation could be a result of interfering backreflections from
the glass cell windows.}. This effect is negligible for laser
powers below 500 W/cm$^2$.

\begin{figure}
\includegraphics[width=3in]{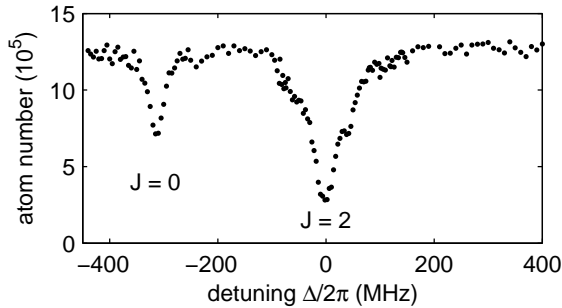}
\caption{Photoassociation spectrum of the excited molecular state
used in the experiment. The two lines belong to the state $| 0_g^-
(\sim S_{1/2}+P_{3/2}), \nu = 1\rangle$ and have rotational
quantum numbers $J=0, 2$ respectively. Shown is the remaining atom
number after exposing a BEC to a 70 $\mu$s light pulse of 460
W/cm$^2$ intensity. The detuning is given relative to the $J=2$
line. Each data point is an average of three measurements. }
 \label{molecularlines}
\end{figure}

In order to identify a suitable molecular level with strong
coupling to the continuum state we investigated molecular lines in
the $1_g$ and $0_g^-$ potentials which connect to the $(S_{1/2} +
P_{3/2})$ and $(S_{1/2} + P_{1/2})$ asymptotes. We choose the
excited state $| 0_g^- (\sim S_{1/2} + P_{3/2}), \nu = 1, J = 2
\rangle$ which is located 26.8 cm$^{-1}$ below the $D$2 line
\cite{Fioretti}. Fig.~\ref{molecularlines} shows the corresponding
photoassociation line together with the line for $J = 0$. At a
laser intensity of 460 W/cm$^2$ the measured atom losses yield a
peak inelastic collision rate $K_\text{inel}=(2 \pm 1) \times
10^{-10} \text{cm$^3$/s}$ which is a factor of 5 weaker than
$K_\text{inel}$ in the example of \cite{Bohn}. Losses due to
excitation of the $D$2 line can be neglected. We observe a strong
intensity dependent light shift of 215 MHz/(kW cm$^{-2}$) of the
photoassociation line which might be mainly explained by coupling
to a $d$-wave shape resonance \cite{Simoni}.

\begin{figure}
\includegraphics[width=3in]{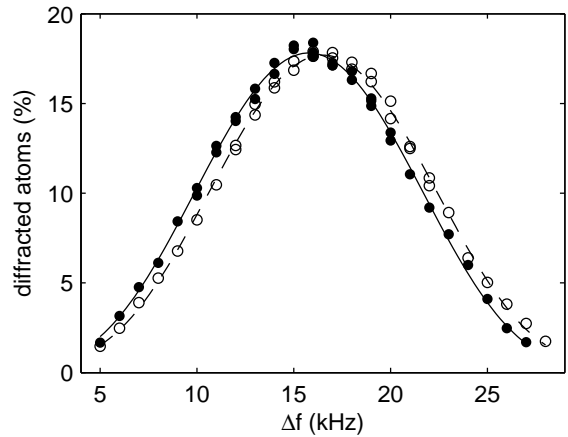}
 \caption{Two Bragg resonance curves with an
optically induced relative shift of 0.75 kHz. The percentage of
the diffracted atoms
 is plotted against the frequency difference of the lattice
beams. The two curves correspond to a detuning $\Delta/2\pi = -
47$ MHz (filled circles) and $\Delta/2\pi = + 47$ MHz (open
circles) at a photoassociation laser intensity of 460 W/cm$^2$.
The lines shown are fits to the data. For better comparison the
right curve (open circles) has been scaled by a factor of 1.09 to
the same height as the left one.}
 \label{braggspectroscopy}
\end{figure}

Measuring the scattering length close to a photoassociation
resonance requires a fast experimental method as atom losses
restrict the observation time to below 100 $\mu$s in our
experiments. Thus the scattering length can neither be extracted
from measurements of the collision rate \cite{Roberts} nor from
the mean-field energy in a condensate expansion \cite{Inouye} both
of which require a few ms. Instead we use Bragg spectroscopy
\cite{Stenger} to determine the mean-field energy by imposing on
the atoms a moving optical lattice composed of two
counter-propagating laser beams with wavenumber $k$ and an
adjustable frequency difference $\Delta f$. The Bragg lattice
transfers a momentum of $2\hbar k$ to the atoms in a first order
diffraction process. This is resonant when energy conservation is
fulfilled, which for {\em noninteracting} atoms reads $ h \Delta
f_0 = (2\hbar k)^2/2m $. For a condensate, however, the resonance
frequency $\Delta f_r$ is shifted by the mean-field energy. In the
Thomas-Fermi approximation this yields a value of
\begin{equation}
  \label{eq:braggspectroscopy}
  \Delta f_r = \Delta f_0 +
            \frac{8 \hbar}{7 m} \ n_0 \ a
\end{equation}
where $ n_0 $ denotes the atomic peak density \cite{Stenger}.
Observing this shift of the Bragg resonance frequency therefore
allows to measure the product of density and scattering length.

We derive the two Bragg beams from a laser which is 1.4 nm blue
detuned relative to the $^{87}$Rb $D$2 line. This determines
$\Delta f_0$ to be 15.14 kHz. Two acousto-optical modulators are
used to control the frequency difference $\Delta f$ between the
two counter-propagating beams. The beams have a diameter of
$\approx 900$ $\mu$m and are aligned along the radial trap axis in
a horizontal direction. In our measurements we apply a 70 $\mu$s
square-pulse of Bragg light to the condensate. After 12 ms of time
of flight, when the momentum components of the condensate have
spatially separated, we use absorption imaging to measure the
portion of condensate atoms that have been diffracted. We always
choose the intensity of the lattice such that about 15\% to 20\%
of the atoms are diffracted at resonance.
 Scanning $\Delta f$ and determining the
percentage of diffracted atoms yields curves as shown in
Fig.~\ref{braggspectroscopy} from which we extract the resonance
positions. Shining in a photoassociation laser pulse (70 $\mu$s
square pulse) at the same time as the Bragg pulse shifts the
resonance position. This shift depends on the detuning $\Delta $
from the molecular line (filled and open circles in
Fig.~\ref{braggspectroscopy}).

For short illumination times $T$ as in our experiment the shape of
the spectra fits well to the Fourier transform of the rectangular
light pulse, $\sin^2 \left(\pi  (\Delta f-\Delta f_r)  T \right)/
(\Delta f-\Delta f_r)^2$, which we use to fit the data (see
Fig.~3). Our measurements show that in spite of the
Fourier-limited width of the Bragg resonance of 13 kHz (FWHM) we
can resolve the peak position to better than $\pm$100 Hz.

When we invert the frequency difference of the Bragg laser beams
and diffract atoms to a momentum state with $-2\hbar k$ instead of
$+2\hbar k$ we notice that the absolute value of the resonance
frequency $|\Delta f_r|$ changes. This
 can be explained by an initial condensate momentum of up to 0.05
$\hbar k$ which we find to slowly vary from day to day. This
initial momentum is due to residual experimental imperfections
like optical dipole forces of a slightly non-centered
photoassociation beam. To eliminate this effect we always measure
$\Delta f_r$ for $+2\hbar k$ as well as for $-2\hbar k$ and then
take the difference.

\begin{figure}
\includegraphics[width=3in]{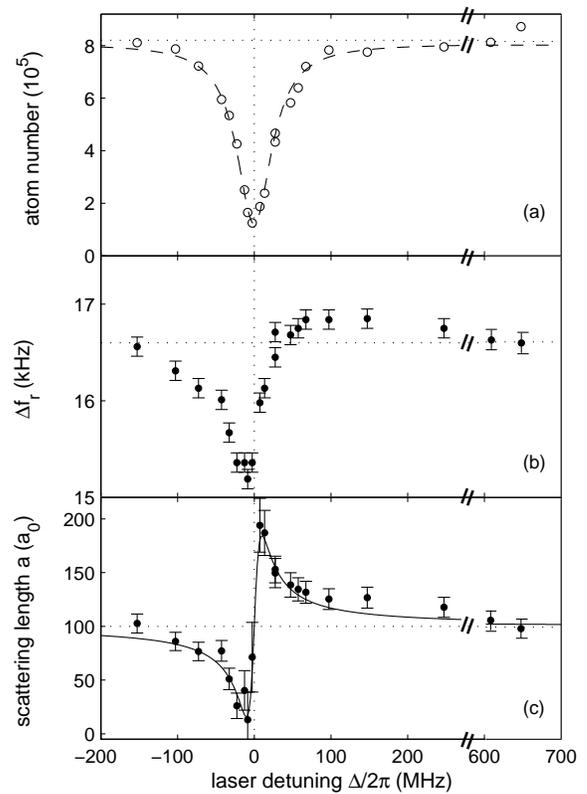}
\caption{Optical Feshbach resonance. In (a) the final atom number
is plotted versus the detuning of the photoassociation laser (the
dashed line is a Lorentz curve to guide the eye). The data in (b)
displays the measured Bragg resonance frequencies. In (c) the
values for the scattering length obtained from the data in (a) and
(b) are plotted.  The continuous line is a fit of
Eq.~(\ref{eq:scatt_length_Julienne}) to the data.}
 \label{scatteringlength}
\end{figure}

Figure \ref{scatteringlength} shows the data we obtain from
scanning the photoassociation laser over the optical resonance for
a fixed laser intensity of 460 W/cm$^2$. The number of atoms in
the condensate at the end of the laser pulse is plotted in
Fig.~\ref{scatteringlength}(a) indicating the position of the
molecular line. On resonance about 90\% of the atoms are lost
after the 70 $ \mu$s of interaction time.
 Fig.~\ref{scatteringlength}(b) shows the resonance frequency $\Delta f_r$ for
Bragg diffraction as a function of laser detuning $\Delta$. For
large positive (and negative) detuning $\Delta $ the value of
$\Delta f_r$ agrees with the 16.6 kHz expected from theory for the
background scattering length $a_\text{bg}=100\,\text{a}_0$
\cite{privTiemann, privJulienneTiesinga} and a BEC with $\approx
8.2 \times 10^5$ atoms. As we tune across the molecular resonance
the measured resonance frequencies exhibit a distorted dispersive
shape. Following Eq.~(\ref{eq:braggspectroscopy}) this is the
result of the combination of two effects: First the scattering
length $a$ varies with $\Delta$ which alone should result in a
dispersive line shape as in Fig.~\ref{theoryplot}. Second the
atomic density $n_0$ decreases due to photoassociation losses
which would, if the scattering length was constant, result in a
symmetrical dip for $\Delta f_r$. On the right hand side of the
resonance these two effects nearly compensate each other whereas
on the left hand side the effects add up to produce a prominent
dip in $\Delta f_r$.

In order to extract the scattering length $a$ from the measured
frequencies one can in a first approach replace the dynamically
changing density $n_0$ in Eq. (\ref{eq:braggspectroscopy}) by a
time averaged value $\langle n_0 \rangle_t$. The average $\langle
n_0 \rangle_t$ can be derived from the  rate equation for the
local density
 \mbox{$\dot{n}=-2K_\text{inel} \
n^2$} \cite{McKenzie}
 describing two-atom losses.
This yields values for $a$ which differ only marginally from the
ones in Fig.~\ref{scatteringlength}(c). The data in
Fig.~\ref{scatteringlength}(c) were obtained from a more detailed
examination which takes into account the full spatially resolved
time evolution of the condensate density \footnote{This
calculation is a simulation of the Rabi flopping between two
levels corresponding to the condensate component at rest and the
component with momentum $2\hbar k$. The changing density due to
the loss is included by introducing a time dependent detuning.}.
This includes the dynamical flattening of the condensate density
profile caused by the rapid atom loss which is much faster than
the trap frequencies \cite{McKenzie}.
Fig.~\ref{scatteringlength}(c) shows that with a laser power of
460 W/cm$^2$ we can tune the scattering length over a range from
10\,$\text{a}_0$ to 190\,$\text{a}_0$. A fit of
Eq.~(\ref{eq:scatt_length_Julienne}) to these data for $a$ yields
a spontaneous decay width $\Gamma_\text{spon}/2\pi $ of 20 MHz and
a resonant inelastic collision rate coefficient $K_\text{inel}=1.7
\times 10^{-10} \text{cm}^3/\text{s}$. These values agree with
those we obtain from atom loss measurements. Thus our data
consistently confirm the intrinsic relation between $a$ and
$K_\text{inel}$  as expressed
 in Eqs. (\ref{eq:scatt_length_Julienne}) and
(\ref{eq:scatt_length_Julienne2}).

 The measured width $\Gamma_\text{spon}/2\pi $ of 20 MHz is larger than the
expected molecular decay width of 12 MHz (corresponding to two
times the atomic width). This might be explained by the line width
of the Ti:Sa laser of about 4 MHz and a power broadening of the
line due to different light shifts of unresolved molecular
hyperfine states \cite{privJulienneTiesinga,McKenzie}.

\begin{figure}
\includegraphics[width=3in]{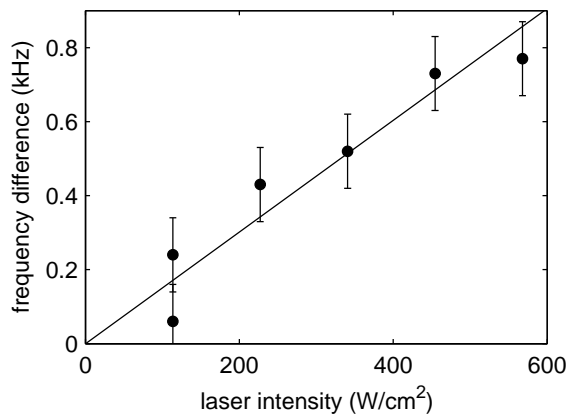}
\caption{Dependence of the optically induced mean-field shift on
the laser intensity.}
 \label{shiftvsintensity}
\end{figure}

Fig.~\ref{shiftvsintensity} demonstrates the linear dependence of
the scattering length $a$ on the photoassociation laser intensity.
For these measurements we determine the Bragg resonance frequency
for the detunings $\Delta/2\pi \approx \pm 50$MHz at various
photoassociation laser intensities.
 This is slightly
complicated by the light shift and broadening of the
photoassociation line which lead to an uncertainty in $\Delta /
2\pi$ of $\pm 10$MHz. We keep the final atom number and density
fixed by adjusting the pulse duration for each laser intensity.
This ensures that only changes in $a$ are reflected in the varying
mean-field shift.  In Fig.~\ref{shiftvsintensity} we plot the
frequency difference $\Delta f(\text{+50\,MHz})-\Delta
f(-\text{50\,MHz)}$ which increases our signal.

In conclusion, our experiments demonstrate the tunability of the
scattering length in ultracold samples by optically coupling free
atoms to a bound molecular state. Because of the exquisite control
one has over laser fields we expect optical Feshbach resonances to
be valuable when it comes to control atom-atom interactions in
demanding applications. The inherent losses suggest the use of
high laser intensities at large detuning and a good choice of the
molecular state in order to optimize the ratio of change in
scattering length and loss rate. Optical Feshbach tuning could be
particularly useful to control atomic interactions in optical
lattices which are discussed as potential future quantum
information processors.

We thank Paul Julienne, Eite Tiesinga, John Bohn, Olivier Dulieu,
Peter Fedichev, Andrea Micheli and Helmut Ritsch for very helpful
discussions. This work was supported by the Austrian Science Fund
(FWF) within SFB 15 (project part 17) and by the European Union in
the frame of the Cold Molecules TMR Network under contract No.
HPRN-CT-2002-00290.

$\dagger$ Permanent address: Department of Physics, Bates College,
Lewiston, ME 04240.

\end{document}